\documentclass[12pt]{article}
\usepackage{epsfig}

\def \be {\begin{equation}}
\def \ee {\end{equation}}
\def \bea {\begin{eqnarray}}
\def \eea {\end{eqnarray}}
\def \nn {\nonumber}

\def \R {{\tt I\kern-.25em R}}
\def \K {{\tt I\kern-.25em K}}
\def \C {{\tt C\kern-.35em C}}
\def \Z {{\tt Z\kern-.3em Z}}
\def \H {{\tt I\kern-.25em H}}
\def \S {{\tt S\kern-.40em S}}
\def \del {\partial}
\def \dels {\partial\kern-.5em / \kern.5em}
\def \As {{A\kern-.5em / \kern.5em}}
\def \Ds {D\kern-.7em / \kern.5em}

\def \a {\alpha}
\def \b {\beta}

\def \eps {\epsilon}

\def \lam {\lambda}

\def \th {\theta}

\def \csch {\mbox{csch}}
\def \RR {{\cal R}}

\begin{document}

\hfill hep-th/0304177\\
\vskip .5in

\begin{center}

\textbf{\bf \large A Note on Acceleration from Product Space
Compactification}

\vskip .5in { Chiang-Mei Chen, Pei-Ming Ho, Ishwaree P.
Neupane, John E. Wang} \vskip 15pt

\sffamily{cmchen,
pmho,
ishwaree,
hllywd2~@phys.ntu.edu.tw
 }\\
\vskip.5in

Department of Physics, National Taiwan University, Taipei 106,
Taiwan, R.O.C.
\\


\vspace{60pt}
\end{center}
\begin{abstract}

We study compactifications of Einstein gravity on product spaces
in vacuum and their acceleration phases.  Scalar potentials for
the dimensionally reduced effective theory are found to be of
exponential form and exact solutions are obtained for a class of
product spaces.  The inflation in our solutions is not sufficient
for the early universe.  We comment on the possibility of
obtaining sufficient inflation by compactification in general.

\end{abstract}
\date{\today}

\setcounter{footnote}{0}

\newpage

\section{Introduction} \label{intro}

In most models of inflation, the source of inflation is a positive
potential energy from a scalar field which dominates over the
kinetic energy.  The precise contribution of the scalar field is
still undetermined, except that the potential has to be flat
enough in order to produce sufficient inflation. Recently it has
been shown that four dimensional accelerating universes can be
obtained from pure gravity in higher dimensions \cite{TW,Ohta,Roy,
Wolf,new}. In these models\footnote{The solutions of
Ref.~\cite{TW} correspond to taking the zero-flux limit of the
S-brane solutions in Refs.~\cite{CM,Ohta1}. Previous work on
accelerating cosmologies from time dependent flat internal spaces
include Refs.~\cite{Cosmo,Cosmo2}.} the role of the inflaton is
played by the scalars coming from the higher dimensional metric
after dimensional reduction. These models therefore have the
attractive feature that the scalar field potential arises in a
natural way and is completely determined by gravity.

The no go theorem in Ref.~\cite{NM}, states that Einstein gravity
coupled to any matter fields with positive kinetic energies can
not generate non-singular warped compactifications to de Sitter
space $dS^d$ (nor $\R^d$) with finite $d$ dimensional Newton
constant, unless there is a positive potential for some scalar
fields.  An assumption in this no go theorem is that the
compactified dimensions are time independent. Discarding this
assumption, Townsend and Wohlfarth \cite{TW} found an accelerating
phase for hyperbolic compactifications\footnote{ We shall assume
that $H_n$ is compact. This can be achieved by taking a quotient
of the noncompact hyperbolic space by a discrete subgroup of the
isometry group. Since the particular choice of the subgroup and
its action does not affect the local Einstein equations, we do not
have to specify these details in this paper. Refs.~\cite{kaloper,
silva} discuss some aspects of compact hyperbolic extra dimensions
for cosmology.} of the form $\R^{3+1}\times H_n$.

The fact that we need negative curvature for acceleration can be
understood from the following observation. For a product space
with metric \be ds^2 = \a^2 ds_4^2 + \b^2 ds_n^2, \ee the Hilbert
Einstein action is  \bea \label{DR}
S &=& \int d^{4+n} x \sqrt{-g}\, \RR \nn \\
&=& \int d^n x \sqrt{g_n} \int d^4 x \sqrt{-g_4} \a^4 \b^n \left[
\frac{ \RR_4 }{\a^2} + \frac{\RR_n}{\b^2} + \cdots  \right]\,,
\eea where $\RR_4$, $g_4$ ($\RR_n$, $g_n$) are the curvature and
metric defined by $ds_4^2$ ($ds_n^2$) \footnote{The additional terms
in the bracket involving derivatives of the scalars $\a$, $\b$
are $$ \frac{- 6 \nabla^2\ln\alpha -
6 (\partial\ln\alpha)^2 + (n^2-n) (\partial\ln\beta)^2}{\a^2}\,.$$
}.

When $\b = \a^{-2/n}$, the $4+n$ dimensional action simplifies
into the sum of the 4D Hilbert Einstein action plus an action for
the scalar field \be \phi = \ln\b, \ee which determines the size
of extra dimensions. The line element $ds^2_4$ is now the
metric in 4D Einstein frame.

The second term in (\ref{DR}) acts as a scalar field potential for
$\phi$
\be \label{V}
V = - \RR_n
e^{-(n+2)\phi}.
\ee
Therefore in order to have a positive potential we need
to compactify a space with negative curvature.  For extra
dimensions of constant curvature, that is spherical or hyperbolic
spaces, we see from (\ref{V}) that the potential has an
exponential form. (For the flat case the potential vanishes.)

Except for very large $n$, the parameters in the theory are all of
order $1$ which is a problem in generating inflation from the
potential in (\ref{V}). To have large acceleration, we need large
$V$, but this also implies large slope $\del V/\del \phi$ and so
the kinetic energy quickly catches up according to the equation of
motion for $\phi$. Therefore the number of e-foldings of such
inflationary models is expected to be only of order $1$. This is
indeed the case for compactification on a single hyperbolic space
\cite{TW}. If we generalize this setup to extra dimensions which
are product spaces for example, this problem might be evaded
because there are more scalars and more complicated potentials.
The situation might be similar to the hybrid model of inflation.
We construct a model in Sec.~\ref{hybrid} showing that sufficient
inflation can be achieved even if all parameters are of order $1$.

The plan of the paper is as follows. In Sec.~\ref{n}, we give a
general formulation of Einstein gravity in vacuum for product
spaces of flat, spherical and hyperbolic spaces. In
Sec.~\ref{1&2}, we begin with a review of compactifications on a
single hyperbolic space $\R^{3+1}\times H_m$, and then analyze the
next simplest case, a product of two compact spaces
$\R^{3+1}\times\K_1\times\K_2$.  Although we did not find
sufficient inflation for these cases, but we show in
Sec.~\ref{hybrid} that in principle, similar models with only
coefficients of order one can produce eternal inflation. In
Sec.~\ref{Product} we begin analysis of general product spaces
$\K_0\times\K_1\times\cdots\K_n$, where the large $d$ dimensions
does not have to be flat and $d$ does not have to be $3$. We have
found exact solutions for the case $\K_1=\K_2=\cdots=\K_n$, where
$\K_i$ can be a spherical, hyperbolic or flat space. Finally we
comment on our results in Sec. \ref{Disc}.

\section{Generic case} \label{n}

In this section we consider Einstein gravity in vacuum for
spacetimes which are products of an arbitrary number of spaces,
which are each either flat, spherical or hyperbolic.  In Subsec.
\ref{EinEq} we first give the vacuum Einstein equations, and then
treating the scale factors as scalar fields we rewrite the theory
from the dimensionally reduced viewpoint in Subsec. \ref{4D}. The
advantage of this approach is that we can now apply our knowledge
of scalar field inflation.

\subsection{Einstein equations for product spaces in vacuum}
\label{EinEq}

Let us consider spacetimes of the form \be \R\times \K_0\times
\K_1\times\cdots \K_n, \ee which are products of flat, spherical
and hyperbolic spaces in addition to the time direction.  Each
factor of $\K_i$ can be either flat, spherical or hyperbolic which
we label by a numbers $\eps_i$, which are $0, 1$ or $-1$,
respectively. The dimension of each space $\K_i$ will be denoted
$m_i$ for $i=0,...,n$ so the total dimension is $\sum_{i=0}^n
m_i+1$.  We define $m_0$ by $d$, and we will find the dimensional
reduction of the higher dimensions to $\K_0$.

Our metric ansatz for the vacuum solution is
\begin{equation}\label{metric1}
ds^2 = - {\rm e}^{2A(t)} dt^2 + \sum_{i=0}^n {\rm e}^{2B_i(t)}
ds^2_i,
\end{equation}
where $ds^2_i = \bar{g}^{(i)}_{ab}dx^a dx^b$
is the metric for the space $\K_{(m_i,\eps_i)}$
given by
\be
ds_i^2 = \left\{
\begin{array}{ll}
d\chi^2 + \chi^2 d\Omega_{m_i-1}^2,
& \mbox{for}\quad \eps_i=0, \\
d\chi^2+\sin^2\chi \, d\Omega_{m_i-1}^2,
& \mbox{for}\quad \eps_i=1, \\
d\chi^2+\sinh^2\chi \, d\Omega_{m_i-1}^2, & \mbox{for}\quad
\eps_i=-1.
\end{array}\right.
\label{dsn} \ee  For each component of the product space, the
Ricci tensor is given by
\begin{equation}
\bar R^{(i)}_{ab} = \epsilon_i (m_i - 1) \bar g^{(i)}_{ab}
\end{equation} so the Ricci tensor for the full metric
(\ref{metric1}) is relatively simple
\begin{eqnarray}
R_{tt} &=& - \sum_{i=0}^n m_i (\ddot B_i + \dot B_i^2 - \dot A
\dot B_i), \\
R^{(i)}_{ab} &=& \left\{ {\rm e}^{2B_i-2A} \left[ \ddot B_i + \dot
B_i \left( - \dot A + \sum_{j=0}^n m_j \dot B_j \right) \right] +
\eps_i (m_i - 1) \right\} \, \bar g^{(i)}_{ab}.
\end{eqnarray}

\ From the above expressions, we see that the equations simplify
using the gauge condition
\begin{equation}\label{gauge}
- A + \sum_{j=0}^n m_j B_j = 0.
\end{equation}
This is merely a gauge condition corresponding to a time
reparametrization since all the functions depend only on time.  In
this gauge, the vacuum Einstein equations reduce to
\begin{eqnarray}
\sum_{i=0}^n m_i (\ddot B_i + \dot B_i^2 - \dot A \dot B_i) &=& 0,
\label{BB}
\\
\ddot B_i + \eps_i (m_i - 1) {\rm e}^{2A-2B_i} &=& 0. \label{B}
\end{eqnarray}
Although the general solutions are not easy to obtain due to the
coupling between the differential equations, we have obtained
exact solutions for some particular cases. We will discuss some
of these solutions below.

\subsection{Effective theory in lower dimension}
\label{4D}

Let us assume that $\K_0$ corresponds to the large spatial dimensions in
which we live, so the $\K_i$'s are extra dimensions. We would like
to dimensionally reduce the theory to $\K_0$ and obtain an
effective theory of $d$ dimensional
gravity coupled to scalar fields.

As noted in Sec.~\ref{intro}, the metric of the $d$ dimensional
Einstein theory is different from the $d$ dimensional part of the
full metric. We rewrite (\ref{metric1}) as \be \label{metric} ds^2 =
\a^2 a^2(-d\eta^2+ds_0^2) +\sum_{i=1}^n a_i^2 ds_i^2, \ee where
$\eta$ is the conformal time in the $d$ dimensional Einstein
frame, $a$ is the $d$ dimensional scale factor and \be
\label{sum1} \a =
\prod_{i=1}^{n} \a_i, \quad \mbox{with} \quad \a_i =
a_i^{-\frac{m_i}{d-1}}. \ee
The Einstein tensor is
\bea
G_{00}&=&\frac{d(d-1)}{2}
\left(\frac{a'}{a}\right)^2 - \rho \, a^2, \\
G_{11}&=&-\frac{(d-1)(d-4)}{2} \left(\frac{a'}{a}\right)^2
-(d-1)\frac{a''}{a} - p \, a^2,
\eea
where $\rho$ and $p$ can
be identified with the energy density and pressure in $d$
dimensional space. The other non-vanishing components of the
Einstein tensor give redundant equations. In the above we have denoted the
derivative of $\eta$ by a prime: $f'=df/d\eta$.

\ From $\rho$ and $p$ we deduce the kinetic and potential
terms\footnote{Potentials coming from compactifications were
discussed in Ref.~\cite{potentials}.} for the scalar fields \bea
K&=&\frac{\rho+p}{2}=
\sum_{i=1}^n\frac{m_i(m_i+d-1)}{2(d-1)a^2}{\phi'_i}^2
+\sum_{i>j=1}^{n}\frac{m_i m_j}{(d-1)a^2}\phi'_i\phi'_j
-\eps_0\frac{d-1}{2a^2}, \nn \\
&=& \sum_{i=1}^n\frac{1}{2a^2}{\varphi'_i}^2
+\sum_{i>j=1}^{n}\frac{2c_i c_j}{a^2}\varphi'_i\varphi'_j
-\eps_0\frac{d-1}{2a^2}, \label{K} \\
V&=&\frac{\rho-p}{2}= \sum_{i=1}^n (-\eps_i)\frac{m_i(m_i-1)}{2}
e^{-\frac2{d-1} \left( (m_i+d-1)\phi_i + \sum_{j\neq i}^{1\leq
j\leq n}m_j\phi_j \right)}
-\eps_0\frac{(d-1)^2}{2a^2}, \nn \\
&=& \sum_{i=1}^n (-\eps_i)\frac{m_i(m_i-1)}{2}
e^{\sqrt{\frac2{d-1}}\left(\varphi_i/c_i +2\sum_{j\neq i}^{1\leq
j\leq n}c_j\varphi_j\right)} -\eps_0\frac{(d-1)^2}{2a^2},
\label{Vphi} \eea where \be c_i = \sqrt{\frac{m_i}{2(m_i+d-1)}},
\ee and the scalar fields $\phi_i$, $\varphi_i$ are defined by \be
a_i=e^{\phi_i}, \qquad \phi_i = -
\frac{\sqrt{2(d-1)}\,c_i}{m_i}\varphi_i. \ee From (\ref{sum1}) we then
have \be
\alpha_i = e^{\sqrt{\frac2{d-1}} \, c_i \varphi_i}. \ee
The last terms in
(\ref{K}) and (\ref{Vphi}) are the contributions from the
curvature of the $d$ dimensional space.

The effective Lagrangian is ($8\pi G_{d+1}=1$)
\footnote{ In our convention, the
factor $\sqrt{-g_4}=a^{d+1}$ is included in $L$, and so the
action is just $S = \int d^{d+1} x L$.}
\be L = \sqrt{-g} \left( \frac{{\cal R}}2 + K - V
\right) = a^{d+1} \left[ -\frac{d(d-1)}{2}\left(
\frac{a'}{a^2}\right)^2
+K-V \right]. \ee
The action for the scalar fields can also be
derived from the Hilbert Einstein action in higher dimensions,
up to total derivatives, as we explain in Appendix A.

It is well known that off-diagonal terms
omitted in (\ref{metric}) are gauge fields from the lower
dimensional viewpoint. We can include their effect by adding
Yang-Mills Lagrangian to the effective action, but the gauge
coupling will be a function of the scalar fields. If we redefine
the gauge field to absorb these factors so that the gauge
coupling is constant, we are effectively introducing charges to
the scalar fields. We will not pursue this effect here.


\section{Simple Product Spaces} \label{1&2}

In this section we study the acceleration of the product spaces
$\R^{3+1}\times H_m$ and $\R^{3+1}\times \K_1\times \K_2$.

\subsection{$\R^{d+1}\times H_m$} \label{1}

Our ansatz for the spacetime metric of $\R^{d+1}\times \K$ is \be
\label{Einstein1} ds^2 = {\alpha}^{2}
\left(-\,a^{2d}\,dt^2+a^2\,dx_{(d)}^2\right)
+{\alpha}^{-\,2(d-1)/m} r_h^2 \,ds_1^2 \, , \ee and $\K$ is taken
to be an $m$-dimensional compact hyperbolic space with the metric
$ds_1^2$ given by (\ref{dsn}) for $\eps_1=-1$.
The ansatz~(\ref{Einstein1}), with \be
\alpha^2(t)=\left(\frac{e^{\,d\lambda_0
t}}{K(t)}\right)^{2m/(m-1)(d-1)}\,, \quad a^2(t)=
\left(\frac{K(t)}{e^{[(m+d-1)/m]\lambda_0 t}}\right)^{2m/(m-1)(d-1)} \,,\ee
solves the vacuum Einstein equations for
 \be \label{SingleH}
 K(t) = \frac{\lambda_0 r_h}{(m-1)}\,\frac{\beta}
 {\sinh\left[\lambda_0\beta |t-t_1|\,\right]}\,,\quad
 \beta=\sqrt{\frac{d(m+d-1)}{m}}\,,\ee
where $t_1$ is an integration constant. This solution
was first found in Ref.~\cite{TW} when $d=3$. Henceforth we take
$\lambda_0=1$, $r_h=1$, and shift the time so that $t_1=0$.

The proper time for a four-dimensional observer (i.e., $d=3$) is
measured by $\tau$, with $d\tau=a^3(t)\,dt$. The
metric~(\ref{Einstein1}) then takes the form \be ds^2 = \a^2(\tau)
 (-d\tau^2+a^2(\tau)\,dx^2) + \a^{-\,4/m}(\tau)ds_1^2\,. \ee According
to (\ref{Vphi}), the potential of $\varphi$ is \be \label{Vm}
V = \frac{m(m-1)}{2}\, e^{\varphi/c_m} \ee
where \be \label{cn} c_m = \sqrt{\frac{m}{2(m+2)}}. \ee
In order to have sufficient inflation, the potential can not be
too steep, otherwise $\varphi$ rolls down the hill too quickly
and the inflationary era will be too short. The slow roll condition
$V^{-1}\frac{dV}{d\varphi} \ll 1$ is satisfied if $c_m \gg 1$.
However, $c_m$ is bounded by $1/2 \leq c_m<\sqrt{1/2}$.


The 4D spacetime is expanding if
\be
\label{expand1}
\frac{da}{d\tau}>0 \quad \Longrightarrow \quad
 n_1(t)\equiv 1+\frac{m\beta}{m+2}\, \coth
 (\beta\,t)<0\,,  \ee
with $\beta$ defined in (\ref{SingleH}). The expansion is
accelerating only if \be \label{accel1}
\frac{d^2a}{d\tau^2}>0\quad \Longrightarrow \quad
n_2(t)\equiv \frac{m(m-1)\beta^2}
 {(m+2)^2\,\sinh^2\left[\beta\,t\right]}> (n_1(t))^2 \,.\ee
 The conditions (\ref{expand1}), (\ref{accel1}) are satisfied
 simultaneously for $t<0$ in a certain interval. A small but
 negative time $t$ actually corresponds to a positive proper
 time $\tau$. Specifically, in the limit $0<t\ll 1$, we find
 \be \tau\sim -\,
\left(\frac{1}{t}\right)^{(m+2)/2(m-1)}\,, \quad a \sim
\left(\frac{1}{t}\right)^{m/2(m-1)} \,. \ee
This also implies that the singularity of the function $K(t)$ at $t=0_+$
corresponds to $\tau=-\infty$ in Einstein frame.

An intuitive understanding of why the acceleration occurs at negative
time $t$ is given in \cite{TW,new}.
Numerical studies show that the number of e-foldings
 \footnote{ The initial time $t_i$ and
final time $t_f$ are defined by the times when the scale factor
$a$ starts and stops to accelerate, respectively. } \be N =
\ln\left(\frac{a(t_f)}{a(t_i)}\right)
=\int_{t_i}^{t_f} dt H \ee is of order $1$, see table $1$.
(Here $H=a^{-1}\,(da/d\tau)$ is the Hubble parameter.)
The solution for the spacetime metric $\R^{3+1}\times H_m$ is
therefore not applicable for implementing inflation
in the early universe.
We note that the initial condition $\varphi'(t_i)$ does not
significantly affect the value of $N$, because a larger initial
velocity will shoot up the hill to a higher point, which is good
for inflation, but it will roll down faster as the slope is also larger
at higher points.


\begin{table}[ht]
\label{table1}
{\bf Table 1: } The period of the accelerated expansion for $d=3$. \\
 \\
\begin{tabular}{|c|c|c|c|} \hline
$~n~$ & $t_i\cong $ & $t_f\cong$ & Ratio
$f=\frac{a(\tau_2)}{a(\tau_1)}$
\\ \hline \hline $2$ & $ -\,0.7367 $ & $-\, 0.1991 $ & $ 1.99$\\ \hline
$3$ & $-\, 0.7249 $ & $ -\,0.1359$ & $2.25$
\\ \hline $4$ & $ -\,0.7259 $ & $-\,0.1051 $ & $2.48 $ \\
\hline $5$ & $-\, 0.7287 $ & $-\, 0.0861 $ & $2.68$ \\
\hline $6$ & $-\, 0.7316 $ & $ -\,0.0731 $ & $2.83 $ \\
\hline $7$ & $-\, 0.7341$ & $ -\,0.0636$ & $3.04 $
\\ \hline
\end{tabular}
\end{table}

\subsection{$\R^{3+1}\times \K_1\times \K_2$} \label{2}

In this section we consider the spacetimes $\R^{3+1}\times
\K_1\times \K_2$, which are the product of a large 3 dimensional flat
space and two compact spaces. The ansatz for the spacetime
metric is
\be ds^2 = \a^2(\eta) a^2(\eta) (-d\eta^2+dx^2)
+\b_1^2(\eta) ds_{\K_1}^2+\b_2^2(\eta) ds^2_{\K_2}, \ee
\bea &\a
= \a_1\a_2, \quad
\a_1 = e^{c_m\varphi_1}, \quad \a_2 = e^{c_n\varphi_2}, \\
&\b_1 = \a_1^{-2/m}, \quad \b_2 = \a_2^{-2/n},
\eea
where the metric for $\K_1$ and $\K_2$ are of the form given in (\ref{dsn}),
and their dimensions are $m$ and $n$.

As shown in Sec.~\ref{n}, the Einstein theory dimensionally reduced
to 4D is equivalent to 4D Einstein gravity coupled to
scalar fields $\varphi_1$, $\varphi_2$ with the kinetic term
\be
K = \frac{1}{a^2} \left(
\frac{1}{2}({\varphi'}_1^2+{\varphi'}_2^2)
+ 2 c_m c_n {\varphi'}_1{\varphi'}_2 \right),
\ee
and the potential
\be
V = V_{m,\,\epsilon_1} e^{2c_n\varphi_2} + V_{n,\,\epsilon_2}
e^{2c_m\varphi_1},
\ee
where $V_{m,\,\epsilon_i}$ are given by
\be
V_{m,\,\epsilon_i}=-\,\epsilon_i\,\frac{m(m-1)}{2}\,e^{\varphi_i/c_m}\,.
\ee

Let us define the new scalars
\be
\psi_1 = \varphi_1+2 c_m c_n\varphi_2, \quad
\psi_2 = c_{mn}\varphi_2,
\ee
so that the kinetic terms take
the canonical form
$K = \frac{1}{2a^2}({\psi'}_1^2+{\psi'}_2^2)$. In these variables the
potential term becomes
\be
V = -\eps_1 \frac{m(m-1)}{2}\, e^{\psi_1/c_m} -
\eps_2 \frac{n(n-1)}{2}\, e^{2c_m\psi_1+c_{mn}\psi_2/c_n},
\ee
\be
c_{mn} = \sqrt{\frac{2(m+n+2)}{(m+2)(n+2)}}.
\ee
This resembles a hybrid model for inflation.

\section{Parameters of order $1$ vs. $60$ e-foldings}
\label{hybrid}

For the spacetimes $\R^{3+1}\times \K_1\times \K_2$ studied in
Sec. \ref{2}, the scalar potential is \be \label{V1} V =
e^{2\psi_1} + e^{\psi_1+\sqrt{3}\psi_2}. \ee for $m=n=2$,
$\eps_1=\eps_2=-1$.  One might be tempted to claim that it is
impossible for this potential to generate sufficient inflation
because all the parameters here are of order one. How can we
obtain 60 e-foldings from such a potential?

Although it is true that for this particular model we
can not obtain sufficient inflation, the argument above
is over-simplifying the problem. Let us demonstrate
this with an explicit example showing that the
interaction between two fields can lead to a
surprising difference from the naive expectation.

\subsection{Numerical results}

The potential we examine
\be \label{V2} V = e^{2\psi_1/3} +
e^{3\psi_1+\sqrt{3}\psi_2}, \ee
is of the same form as the
one in (\ref{V1}), where all parameters are of order one. The only
difference between these two potentials is the
coefficients of $\psi_1$ in the exponents.

The equations of motion for the scalar fields
expressed in the physical time $t$ ($dt = ad\eta$) are
\be
\ddot{\psi}_i+3H\dot{\psi}_i+\frac{\del V}{\del\psi_i} = 0,
\ee
and the Friedman equation
\be
H^2 = \frac{1}{3}(K+V)
\ee
determines the evolution of the scale factor $a(t)$.

We have studied the expansion of this  model numerically. For two
2nd order differential equation, we can set four initial values
$\psi_1, \psi_2, \dot{\psi}_1$ and $\dot{\psi}_2$. If we fix the
initial condition so that it coincides with the starting point of
acceleration, we first set three of the initial values,
$\psi_1(0)$, $\dot{\psi}_1(0)$, $\dot{\psi_2}(0)$, and this fixes
the initial value of $\psi_2(0)$ by requiring that $\ddot{a} = 0$,
or equivalently $V=2K$ at $t=0$. A set of such initial conditions
is $\psi_1(0)=0, \psi_2(0)=0, \dot{\psi}_1=1, \dot{\psi}_2=1$.
Note that all the initial values of this set are of order $1$, and
the results will be qualitatively the same for variations of order
$1$ for the initial conditions.

\begin{figure}[tp]
\begin{minipage}{70mm}
\begin{center}
\includegraphics*[width=6cm]{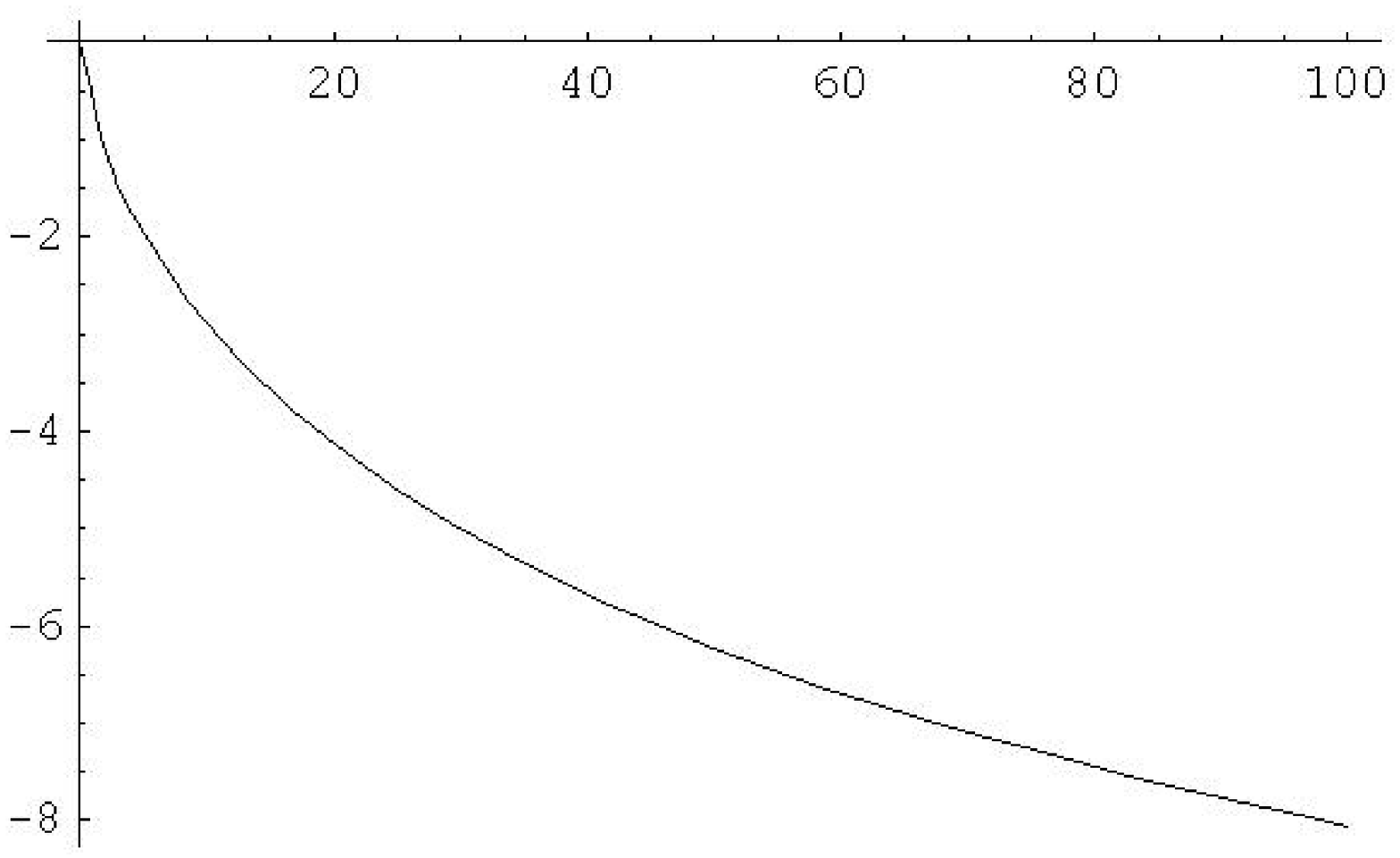}
\centerline{ \hspace{.3in} Figure 1: The evolution of $\psi_1$
for} \centerline{ \hspace{.3in} the potential (\ref{V2}).}
\label{hybrid1psi1}
\end{center}
\end{minipage}
\hspace*{12mm}
\begin{minipage}{70mm}
\begin{center}
\includegraphics[width=6cm]{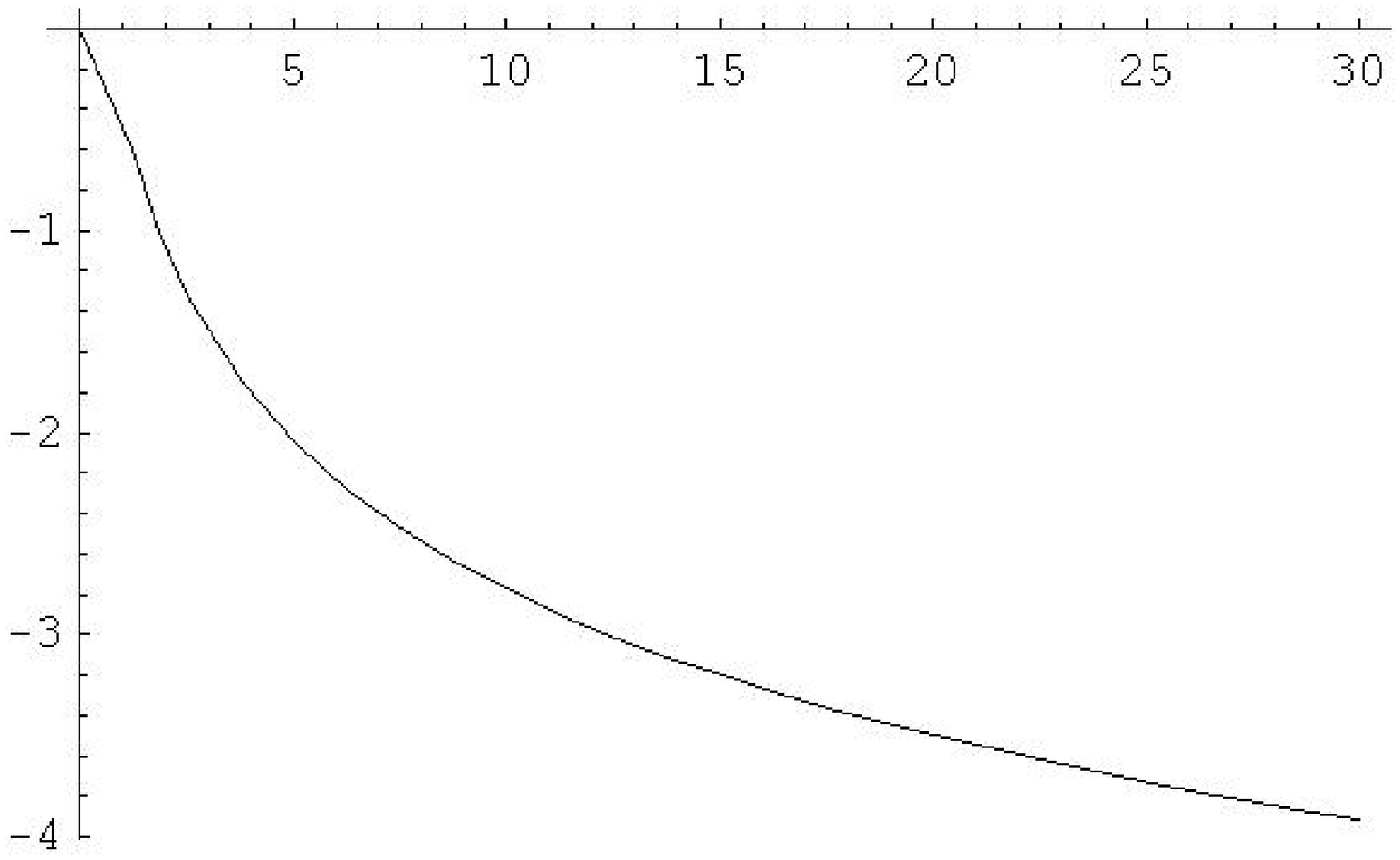}
\centerline{Figure 4: The evolution of $\psi_1$ for} \centerline{
the potential (\ref{V1}). } \label{hybrid0psi1}
\end{center}
\end{minipage}
\begin{minipage}{70mm}
\begin{center}
\includegraphics[width=6cm]{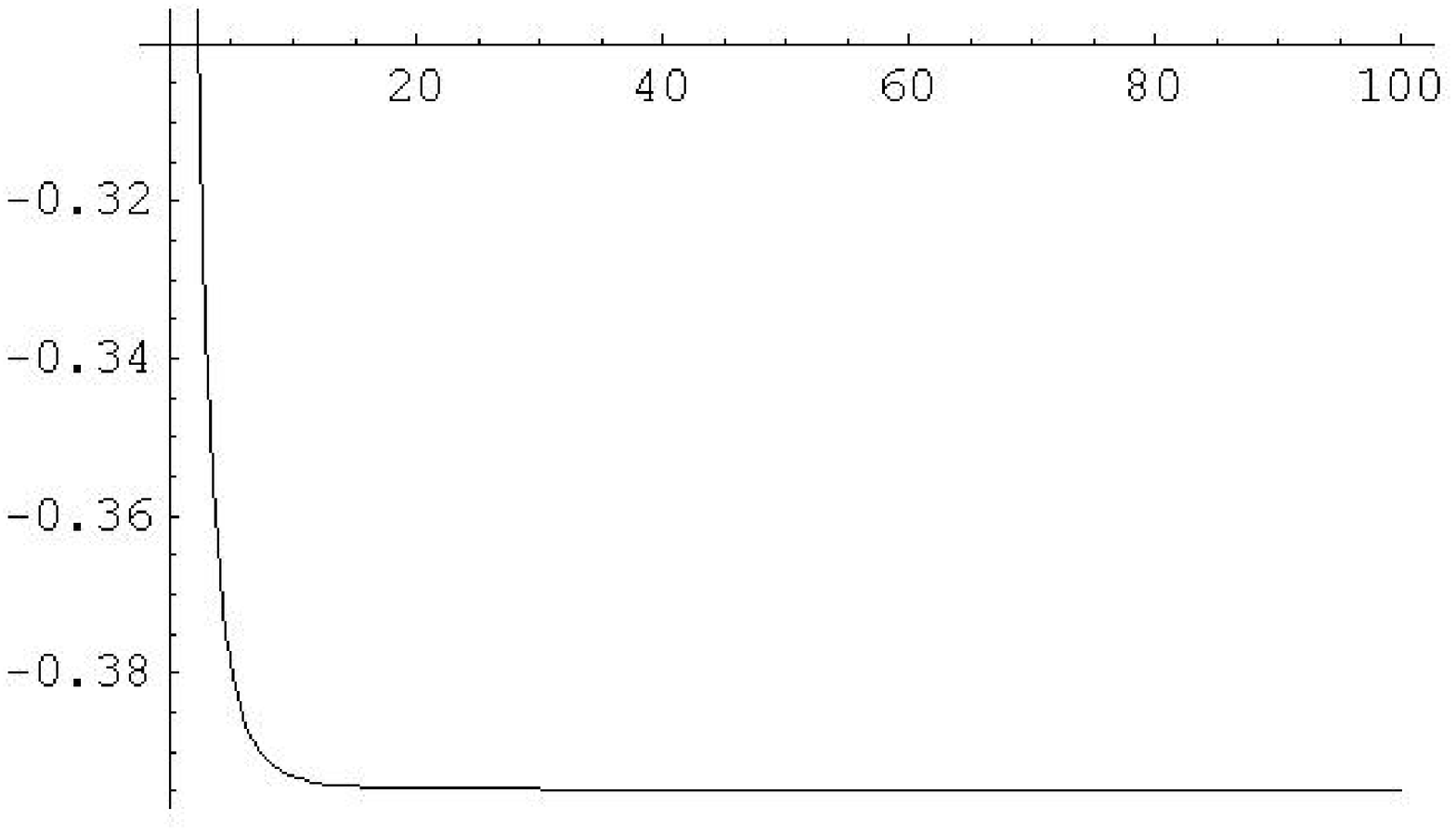}
\centerline{Figure 2: The evolution of $\psi_2$ for} \centerline{
the potential (\ref{V2}). } \label{hybrid1psi2}
\end{center}
\end{minipage}
\hspace*{12mm}
\begin{minipage}{70mm}
\begin{center}
\includegraphics[width=6cm]{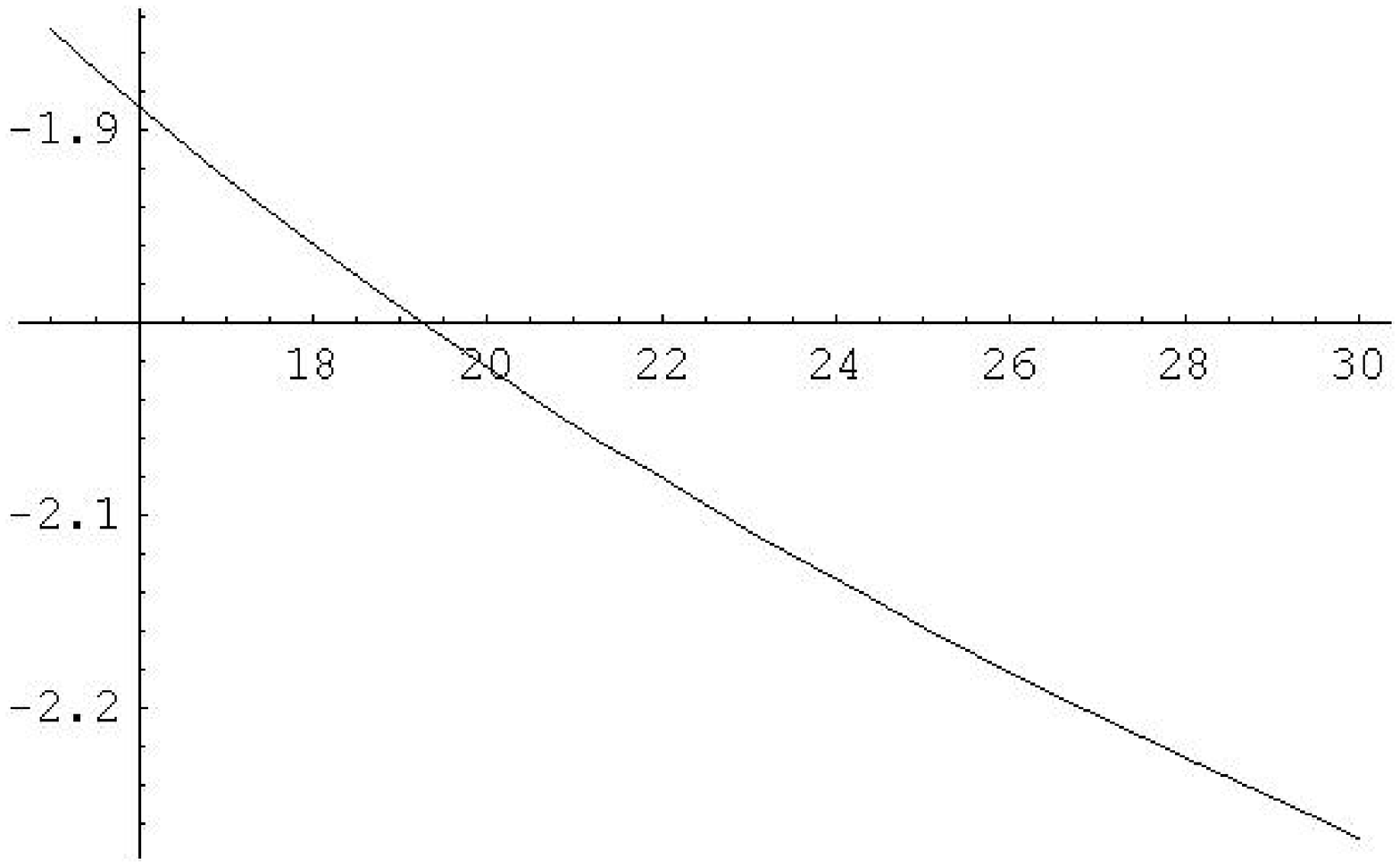}
\centerline{Figure 5: The evolution of $\psi_2$ for}
\centerline{the potential (\ref{V1}). } \label{hybrid0psi2}
\end{center}
\end{minipage}
\begin{minipage}{70mm}
\begin{center}
\includegraphics[width=6cm]{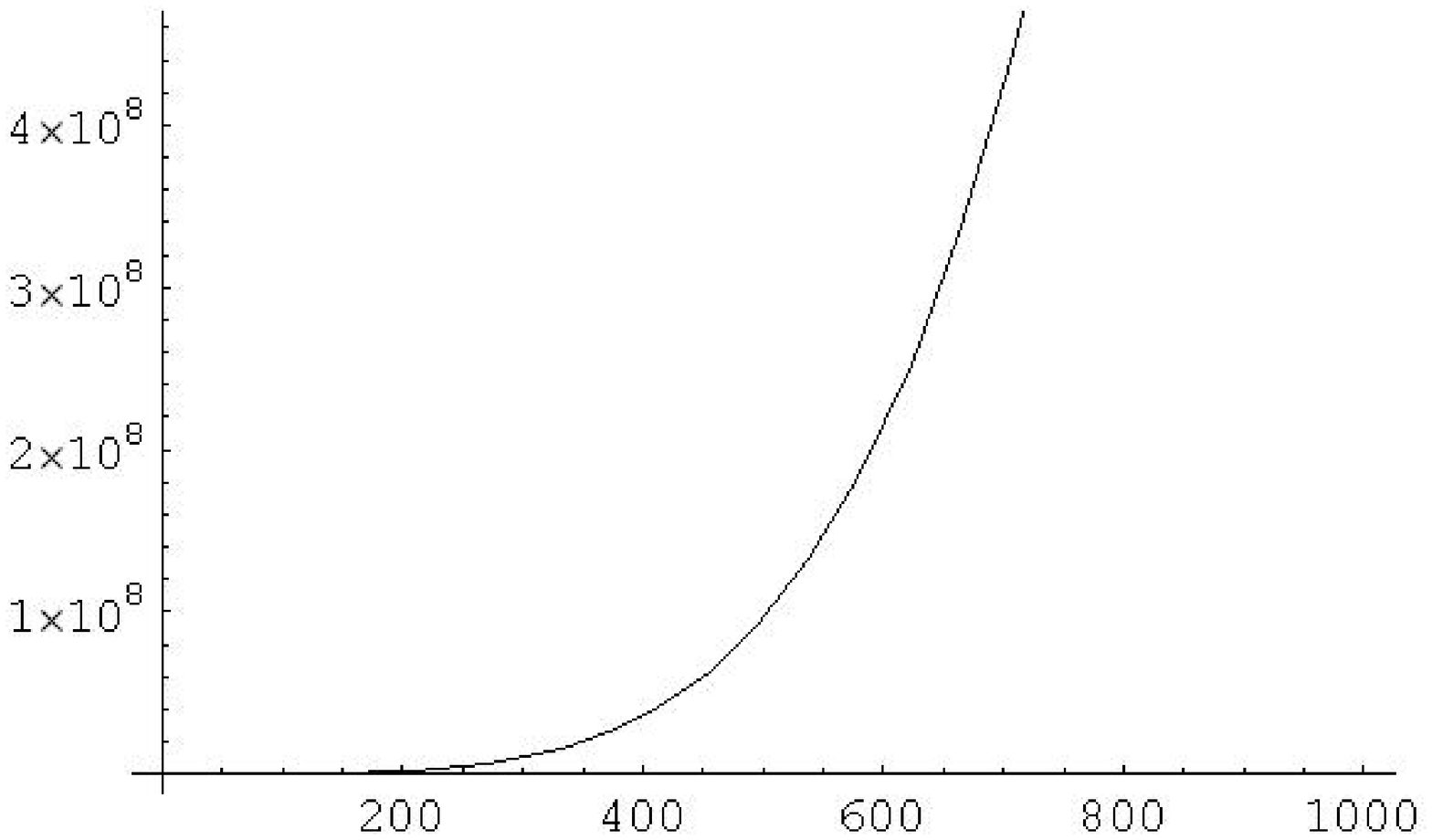}
\centerline{Figure 3: Evolution of the scale factor} \centerline{
$a(t)$ for the potential (\ref{V2}).} \centerline{The acceleration
never stops.}
 \label{hybrid1a}
\end{center}
\end{minipage}
\hspace*{12mm}
\begin{minipage}{70mm}
\begin{center}
\includegraphics[width=6cm]{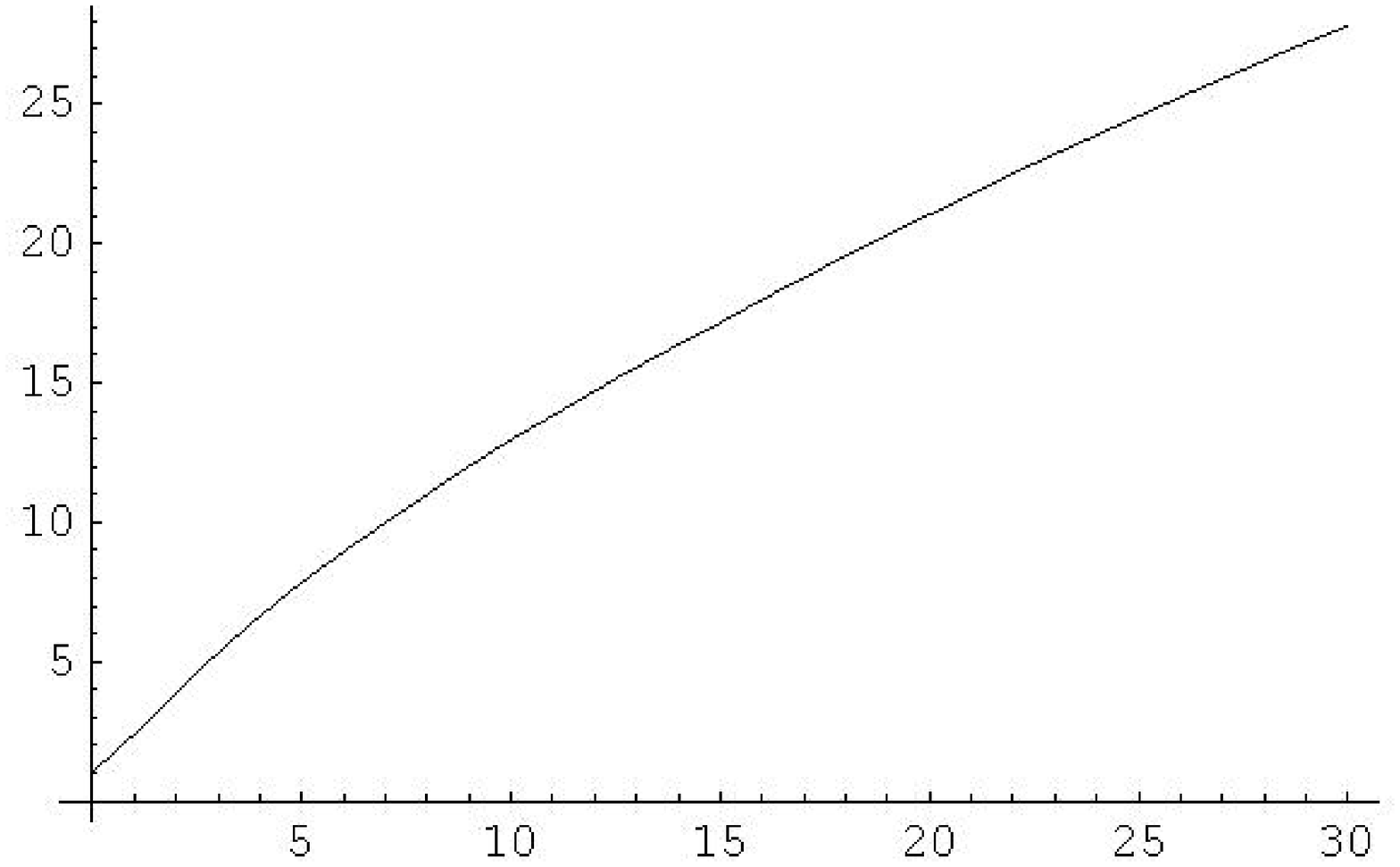}
\centerline{Figure 6: Evolution of the scale factor }
\centerline{$a(t)$ for the potential (\ref{V1}).} \centerline{
Acceleration for a very short}  \centerline{period of order $1$
occurs near} \centerline{ the origin $t=0$. } \label{hybrid0a}
\end{center}
\end{minipage}
\end{figure}

Our numerical study showed that the acceleration never stops, but
asymptotes to zero acceleration, $\ddot{a} \rightarrow 0$. The
evolution of $\psi_1$, $\psi_2$ and the scale factor $a(t)$ are
given in Figures 1, 2 and
3, respectively. In contrast, the results for the
potential (\ref{V1}) from dimensional reduction are plotted in
Figures 4-6. Note that the time
scales of the plots are quite different.

The Hubble parameter $H = \dot{a}/a$
is found to be $0.004, 0.0004, 0.00004$
at $T=10^3, 10^4, 10^5$, respectively.
The corresponding e-folding number is
$21, 32, 42$, respectively.
It is remarkable that while the two potentials
(\ref{V1}) and (\ref{V2}) are so similar,
their spacetime evolutions are so drastically different.

The behavior of the coupled system is quite interesting. We see
from the figure of $\psi_2$ that it asymptotes to a constant as
$t\rightarrow \infty$. The only difference we noticed between the potentials
(\ref{V1}) and (\ref{V2}) which might be the crucial here is that
the relative magnitude of the coefficients of $\psi_1$ in the two
exponents in $V$ is reversed. It will be very interesting to
understand better the mechanism behind the model (\ref{V2}) that
leads to eternal inflation. This knowledge will be important for
inflationary model building.

\subsection{Exact solutions}

In the above we studied numerically models of two scalar fields
with the exponential potential \be V =
e^{b\psi_1}+e^{c\psi_1+d\psi_2}. \ee In fact, exact solutions can
also be obtained. From the ansatz \be \label{ansatz} \psi_i = p_i
\ln \tau + q_i, \quad H = \frac{h}{\tau}, \ee we find the solution \bea
&p_1=-\frac{2}{b}, \quad
p_2=-\frac{2(b-c)}{bd}, \\
&e^{bq_1}=\frac{2(3h-1)(d^2-c(b-c))}{b^2 d^2}, \quad
e^{cq_1+dq_2}=\frac{2(3h-1)(b-c)}{bd^2}, \eea and \be h =
\frac{2(d^2+(b-c)^2)}{b^2 d^2}. \ee

These special solutions are either acceleration for all $\tau$ or
deceleration for all $\tau$. The condition for acceleration is $h>1$.
For the potentials from compactification on $\K_1\times\K_2$ in
Sec \ref{2}, \be h=\frac{m_1+m_2}{m_1+m_2+2} < 1, \ee and there is
no inflation. For the hypothetical model (\ref{V2}) we studied
numerically in the previous subsection, $h=38/3>1$.

To understand the form and simplicity of these solutions it is
useful to change variables from $\psi$ to $\varphi$. For the case
where the extra dimensions are $H_2\times H_2$, we have \bea
b=2, \quad c=1, \quad d=\sqrt{3}, \\
p_1=-1, \quad p_2=-\frac{1}{\sqrt{3}}. \eea
The time dependence of the $\varphi$'s is given by
\begin{equation}
\varphi_1=
-\frac{2}{3} \ln \tau + \mbox{constant}, \quad
\varphi_2=
-\frac{2}{3} \ln \tau + \mbox{constant},
\end{equation}
so we find $\varphi_1$ and $\varphi_2$ have the same time
dependence. This result is consistent with the fact that
$\varphi_1=\varphi_2$ is a symmetry axis of the potential; our
numerical simulations confirm the intuition that most solutions
tend to fall along this line of attraction due to energy lost to
cosmic friction.

Comparing these solutions to the general solution for a single
product space $H_m$, we see that these exact solutions can be
understood as the large $\tau$ limit of the generic solutions
\cite{new}.  If we understand these solutions as representing
future infinity, we see that although we can start with generic
solutions, their eventual fate seems to be that the different
hyperbolic spaces attain the same time dependence.

In the next section we will find exact solutions where the product
spaces all have the same time dependence and we find that
inflation is not improved. If all solutions are generically
attracted to the symmetry axis, then eternal inflation is
apparently ruled out when the product spaces are all of the same
type.

Apparently the ansatz (\ref{ansatz}) can be easily applied to an
arbitrary number of scalar fields with exponential potential.
Exact solutions of this class will be reported in a follow up
paper.

\section{Product spaces} \label{Product}

The eternal inflation in the hybrid model of
the previous section motivates us to look for
solutions for more general product spaces as
extra dimensions.

\subsection{Equal Product Spaces}

In this subsection, we discuss the case of $n$ copies of $\K_i$: $\eps_0=0,
m_0=d, m_1=\cdots=m_n=m, B_1=\cdots=B_n=B$ and
$\eps_1=\cdots=\eps_n=\eps$. For this case the equation for
$B_0$ can be solved
\begin{equation}
B_0 = \alpha_0 t + \beta_0,
\end{equation}
where $\alpha_0$ and $\beta_0$ are integration constants, and the
variable
\be
A = d (\alpha_0 t + \beta_0) + m n B
\ee
is determined by the gauge condition (\ref{gauge}). Then
the only equations we need to solve are (\ref{BB}) and (\ref{B}) for $B$
\begin{eqnarray}
\alpha_0^2 d (1-d) + m n [ \ddot B + (1-m n) \dot B^2 - 2 \alpha_0 d
\dot B ] &=& 0, \label{Heq1} \\
\ddot B + \eps (m - 1) {\rm e}^{2d(\alpha_0 t + \beta_0) + 2(mn-1)B}
&=& 0. \label{Heq2}
\end{eqnarray}
The change of variables
\begin{equation}
B = \sqrt{\frac{m-1}{mn-1}} f - \frac{d}{mn-1} (\alpha_0 t + \beta_0),
\end{equation}
\begin{equation}
A = mn \sqrt{\frac{m-1}{mn-1}} f - \frac{d}{mn-1} (\alpha_0 t +
\beta_0).
\end{equation}
simplifies the equations of motion
\begin{equation}
\ddot f + \eps \lambda {\rm e}^{2 \lambda f} = 0, \qquad
\leftrightarrow \qquad \dot f^2 + \eps {\rm e}^{2 \lambda f} =
f_0^2,
\end{equation}
where $\lambda=\sqrt{(m-1)(mn-1)}$. The solutions of this differential
equation are
\begin{equation}
f(t) = \left\{ \begin{array}{ll}
 \frac1{\lambda}\ln\left(\frac{f_0}{\sinh[\lambda f_0 |(t-t_1)|
   ]} \right) \qquad & \eps = -1. \\
 f_0 (t-t_1) & \eps = 0. \\
 \frac1{\lambda}\ln\left(\frac{f_0}{\cosh[\lambda f_0 (t-t_1)
   ]} \right), & \eps = +1. \end{array} \right.
\end{equation}
where the value of $f_0$ is fixed from (\ref{Heq1}) to be
\begin{equation}
f_0 = \sqrt{\frac{(mn+d-1) d}{mn(m-1)(mn-1)}} \, \alpha_0.
\end{equation}
The constants $\alpha_0$ and $\beta_0$ can be set as $\alpha_0=1$ and
$\beta_0=0$ by rescaling and shifting time $t$.

Following the ansatz (\ref{metric1}), the ($d+1$)-dimensional
Einstein frame metric for this generic case is
\begin{equation}
ds_{d+1}^2 = {\rm e}^{\frac{2mn}{d-1}B} \left( - {\rm e}^{2A}
dt^2 + {\rm e}^{2B_0} dx_d^2 \right)\,.
\end{equation}
When $\K$ is a hyperbolic space (i.e., $\epsilon=-1$), we have
\begin{equation} \label{hypermetric1}
ds_{d+1}^2 = - a^{2d} dt^2 + a^2 d x^2,
\end{equation}
with
\begin{equation}
a(t) = h_1(t) h_2(t), \quad
h_1(t) = e^{ - \frac{mn+d-1}{(d-1)(mn-1)} t }, \qquad
h_2(t) = \left( \frac{f_0}{\sinh[\lambda f_0 |(t-t_1)|]}
\right)^{\frac{mn}{(d-1)(mn-1)}}.
\end{equation}
In terms of the proper time coordinate $\tau$, with
\begin{equation}
d\tau = a^d d t,
\end{equation}
the metric~(\ref{hypermetric1}) can be written in the standard
FLRW form as
\begin{equation}
ds_{d+1}^2 = - d\tau^2 + a^2(\tau) dx_d^2.
\end{equation}
The accelerating phase occurs when
\begin{equation}
\del_\tau a > 0, \quad
\partial_\tau \partial_\tau a > 0,
\end{equation}
which, after a straightforward calculation, give
\bea
&- \sqrt{mnd}\coth\th - \sqrt{mn+d-1} > 0, \\
&d(mn-1)\csch^2\th - (\sqrt{mnd}\coth\th + \sqrt{mn+d-1})^2 >0,
\eea where $\th = \lam f_0 (t-t_1)$.  It is interesting that this
equation is dependent only on the product $mn$ and not $m$ or $n$
separately.  For these particular solutions we find that the
period of acceleration and the number of e-foldings are the same
for $H_{mn}$ and product spaces such as $n$ copies of $H_m$.

Analysis of these conditions shows that acceleration phases exist
for all values of $d\geq 1, m \geq 2, n\geq 1$. We find larger
e-foldings for smaller $d$ and larger $mn$. For $d=1$, inflation
can last forever, although the acceleration eventually approaches
to zero. For $d=3$, the number of e-foldings is of order one for
$mn$ of order 100 or smaller.

We finally note that for the case of a simple product space, going
to higher dimensions $d$ and $m$ actually decreases the amount of
inflation. In arbitrary $d$ dimensions, the slow roll condition
becomes
\begin{equation}
\left( \frac{V^\prime}{V} \right)^2 \ll \frac{d}{(d-1)^2}
\end{equation}
which is consistent with our results that inflation for $d=1$ can
last forever.  For the spacetime $R^{1,d}\times H_m$ the potential
is of exponential form and we find
\begin{equation}
\left( \frac{V^\prime}{V} \right)^2 = 4\ \frac{m+d-1}{m(d-1)}
\end{equation}
which is of order $1/d$ for the case where we take $m\approx d$,
and also for the case where we take $m$ large. While going to
large $d$ does flatten out the potential which is good for
obtaining large inflation, going to large $d$ also increases the
cosmic friction. Going to large $d$ therefore does not help us
satisfy the slow roll conditions and does not increase the amount
of inflation.

Apparently there is no room for a realistic inflation model for
the early universe using the solutions found in this subsection.

\section{Discussion} \label{Disc}

In this paper, we generalized the work of \cite{TW}
and considered compactifications of product spaces
of flat, spherical and hyperbolic spaces.
Unfortunately, for all the cases we studied,
the acceleration phases are still not sufficient.
Nevertheless,
we can not yet conclude that
$60$ e-foldings cannot be achieved from pure gravity
compactification.

For the product spaces considered in Sec.~\ref{Product},
the exact solutions we found are only very special solutions.
In general each space in the product can have
a different scale factor,
but we have only found solutions with
a single overall scale factor
for the extra dimensions.
At least numerically,
one can also study products spaces
for which we did not obtain exact solutions.
In view of the importance of interactions between
scalar fields which we discussed in Sec.~\ref{hybrid},
the phase diagram for the solutions can be very complicated
and the implication of the special solutions
can not be taken too far.

Since we know that radiation induces deceleration,
adding off-diagonal components into the full spacetime metric
does not seem so promising. Similarly, in view of the no go
theorem of \cite{NM}, adding matter fields will probably not
help. It would be interesting to know whether
modifications to Einstein theory, such as
Brans Dicke theory or adding Gauss Bonnet terms
will admit solutions with sufficient acceleration.

If we assume that the compactified
space is stabilized before nucleosynthesis
in order not to violate observational data,
it will be hard to apply the acceleration
associated with the shrinking of extra dimensions
to explain the acceleration of the present universe.
This is why we have kept in mind the inflation
of the early universe as our main application.
But a more careful study is needed before this
possibility is completely ruled out.

Although this original setup of Townsend and Wohlfarth does not
provide large e-foldings for a three dimensional universe, it is
possible that with modification, solutions with large amounts of
inflation can be found from a suitable high energy starting point
using gravity as the source of inflation. In a follow up paper we
will discuss the details of these more complicated scenarios.

\bigskip

{\bf {\large Acknowledgements}}:  We thank Je-an Gu for useful
discussions. This work is supported in part by the National
Science Council, the Center for Theoretical Physics at National
Taiwan University, the National Center for Theoretical Sciences,
and the CosPA project of the Ministry of Education, Taiwan, R.O.C.
The work of P.\ M.\ H. is also supported in part by the Wu
Ta-Yu Memorial Award of NSC.

\begin{appendix}
\section{Dimensional reduction}
Consider a product space metric
\begin{equation}
ds^2 = G_{MN} dx^M dx^N = e^{2\phi} g_{\mu\nu} dx^\mu dx^\nu +
\sum_{i=1}^n e^{2\phi_i} ds_i^2, \qquad \mu=0,...,d,
\end{equation}
where each $ds_i^2$ characterized by $\epsilon_i$ can be flat
($\epsilon_i=0$), spherical ($\epsilon_i=1$) or hyperbolic space
($\epsilon_i=-1$) with dimensions $m_i$. The total
spacetime dimension is $D = d + 1 + \sum_{i=1}^n m_i$.
For this ansatz, the $D$-dimensional
Hilbert-Einstein action decomposes as
\begin{eqnarray}
S &=& \int d^D x \sqrt{-G} \, {\cal R}_G, \nonumber \\
&=& \int d^m x \int d^{d+1} x \sqrt{-g} \, \exp\left[(d-1)\phi +
\sum_{i=1}^n m_i \phi_i \right] \, \Bigg\{ {\cal R}_g - 2 d
\nabla^2 \phi \nonumber\\
&-& d (d-1) (\partial \phi)^2 + \sum_{i,j=1}^n m_i m_j \,
g^{\mu\nu} \partial_\mu \phi_i \partial_\nu \phi_j - \sum_{i=1}^n
m_i (\partial \phi_i)^2 \nonumber\\
&+& \sum_{i=1}^n \epsilon_i m_i(m_i-1) e^{2\phi-2\phi_i}
\Bigg\}.
\end{eqnarray}
If we go to Einstein frame using
\begin{equation}
(d-1) \phi + \sum_{i=1}^n m_i \phi_i = 0,
\end{equation}
then the term $-2d\sqrt{-g}\,\nabla^2\phi$ becomes a total
derivative which can be neglected. Finally, the
$d$-dimensional effective action becomes
\begin{eqnarray}
S &=& \int d^{d+1} x \sqrt{-g} \, \Bigg\{ {\cal R}_g - \frac1{d-1}
\sum_{i,j=1}^n m_i m_j g^{\mu\nu} \partial_\mu \phi_i
\partial_\nu \phi_j - \sum_{i=1}^n m_i (\partial \phi_i)^2
\nonumber\\
&+& \sum_{i=1}^n \epsilon_i m_i(m_i-1) \, \exp\bigg( -
\frac2{d-1} \sum_{j=1}^n m_j \phi_j - 2 \phi_i \bigg) \Bigg\}.
\end{eqnarray}

\end{appendix}

\end{document}